\documentclass[a4paper,12pt]{article}
\usepackage{amsmath}
\usepackage{amssymb}
\usepackage{amsfonts}
\usepackage{graphics}
\usepackage{epsfig}

\begin{document}

\renewcommand{\theequation}{\thesection.\arabic{equation}}
\thispagestyle{empty}
\vspace*{-1.5cm}

\begin{center}
{\Large\bf The Generalized Curvature and Christoffel Symbols \\
 for a Higher Spin Potential in $AdS_{d+1}$ Space}\\
\vspace{4cm}
{\large Ruben Manvelyan ${}^{\dag\ddag}$ and Werner R\"uhl
${}^{\dag}$}
\medskip

${}^{\dag}${\small\it Department of Physics\\ Erwin Schr\"odinger Stra\ss e \\
Technical University of Kaiserslautern, Postfach 3049}\\
{\small\it 67653
Kaiserslautern, Germany}\\
\medskip
${}^{\ddag}${\small\it Yerevan Physics Institute\\ Alikhanian Br.
Str.
2, 0036 Yerevan, Armenia}\\
\medskip
{\small\tt manvel,ruehl@physik.uni-kl.de}
\vspace{3cm}
\begin{abstract}
The  generalized curvature tensor and Christoffel symbols are determined in
$AdS_{d+1}$ background  by a modified ansatz of the de Wit - Freedman type by
imposing gauge invariance. The resulting set of recurrence relations and difference equations is solved.
The Riemann curvature tensor is derived by antisymmetrization. All results are presented as
finite power series in the inverse $AdS$ radius and are unique. The fourth order, which is complete for
fields up to spin five, is calculated explicitly. Higher orders can be obtained with the same method.
\end{abstract}
\vspace{4cm}
{\it July 2007}
\end{center}
\newpage
\section*{Introduction}
\quad In this paper we present a recursive procedure to construct generalized Christoffel symbols and the curvature for higher spin (HS) gauge potential in $AdS_{d+1}$ background.
The linearized HS gauge field's geometry on flat space was constructed by B.~de Wit and D.Z.~Freedman in the seminal article \cite{DF} and for spin $s=3$  in \cite{Damour}. The authors presented there a very elegant geometrical hierarchy of generalized Christoffel symbols defined from the gauge transformation properties in a flat background. This construction is still very important and interesting for many reasons. First of all this hierarchy is a full geometrical partner of  Fronsdal's linearized dynamics \cite{F} for  HS gauge fields in the case of flat space and puts this theory on the same level as linearized gravity. From another hand this hierarchy should be intimately connected with the  generalized Weyl tensors introduced in \cite{LV} and with the recent investigations of the so-called unfolded formulation for higher spin theories (see \cite{Vas} and refs. there) which is pretending today to describe the nonlinear interaction for HS fields in a general background. The $AdS$ background for HS fields was always necessary to construct a consistent HS interaction with gravity \cite{Vas1} but became especially important after discovery of the  $AdS_{4}/CFT_{3}$ correspondence of the critical $O(N)$ sigma model and four dimensional HS gauge theory in anti-de Sitter space \cite{Kleb}. This proposal increased the interest in the general problems of the quantum interacting HS  theories in $AdS$ space.

Our interest in the de Wit-Freedman construction in $AdS$ background originates from our last results for the trace anomaly of the conformal scalar in the external HS gauge field in the bulk of $AdS_{4}$ obtained in \cite{MR1} and \cite{MR2}. Our main task is to express and explain our one-loop quantum results in terms of geometrical objects like the generalized Weyl tensor and the generalized Euler density. This we hope to materialize in the next publication applying the results obtained in this article as a first but sufficiently complicated and nontrivial technical step.
So here we present only the unique construction for the HS curvature and Christoffel symbols in $AdS_{d+1}$ background.

The aim is to construct the generalized curvature and Christoffel symbols linearized in the HS field from an ansatz
presenting a $1/L^{2}$ polynomial expansion with $L$ the $AdS$ radius, whose leading term corresponding to $L=\infty$
is the de Wit-Freedman expression \cite{DF}. The terms of higher order $k$ in $1/L^{2}$ contain numerical coefficients  $A_{r_1r_2r_3}^{(l)}, r_1+r_2+r_3 = k$ which we intend to determine by imposing gauge invariance. This necessitates the derivation of a system of equations which is linear in the $A$s. This system is based on a matrix of triangular shape
with blocks of difference operators of first order on the diagonal. Solving this system of equations is the main part of our analysis. At the order $k=2$ the essential problems of this issue all occur and we solved them. So we are
convinced that our methos works to any order up to the highest, $k=[s/2]$, where $s$ is the spin of the HS field.

In the first Section we review and reformulate in our compressed notations the flat space definitions of \cite{DF}  and correct the important formula obtained in
\cite{Ber1} deriving the corresponding normalization factor in Appendix A. In Section 2 we develop the Lie algebra of
covariant derivatives on symmetric tensor fields in $AdS$ space obtaining important commutation formulas. These formulas we apply in Section 3 to calculate the gauge variation of the modified de Wit-Freedman type ansatz in all orders of the $AdS$ curvature and to derive the matrix and the inhomogeneity of the system of equations. In Section 4 we develop the concept of obstruction and present a full solution for the recursion relations to first and second order (Appendix B) in the polynomial expansion in powers of the $AdS$ inverse radius squared. In Section 5 we derive the Riemann curvature tensor from the de Wit-Freedman expression by applying an antisymmetrizer which renders the Riemann tensor the appropriate symmetry of a Young diagram with two rows of equal length $s$. This method has been presented in Section 1 for the flat space. The whole article is written in a mathematical language which represents symmetric tensor
fields as homogeneous polynomials in vectors of the tangent space of degree equal the spin of the field. This elementary language was always used in quantum mechanics (e.g. for electromagnetic multipoles) and in the last thirty years commonly used in conformal field theory for spaces of arbitrary dimensions.

\section{The basic definitions on flat space}
In analogy with the electromagnetic potential $A$ of spin one which possesses
the field tensor $F$ as a gauge invariant curvature, we consider higher spin potentials $h^{(s)}$ as symmetric tensors of rank $s$ and their gauge invariant curvature $R$ of tensor rank $2s$. This curvature is the simplest gauge invariant object that is linear in the potential, is obtained from the potential
by acting on it with a differential operator, but does not vanish if the free field equations are applied. These equations are
\begin{equation}
h^{(s)\lambda \nu}_{\lambda \nu \mu_{1} \mu_{2}...\mu_{s-4}} = 0
\label{1.1}
\end{equation}
which is the postulate of vanishing double trace, and a second order field equation. To formulate it we introduce the concise notation of homogeneous functions for symmetric tensors, namely we contract these tensors with vectors from the tangent space denoted $a$, $b$ and similarly
\begin{equation}
h^{(s)}(z; a) = \left(\prod_{i=1}^{s} a^{\mu_i}\right) h^{(s)}_{\mu_1 \mu_2...\mu_s}(z)
\label{1.2}
\end{equation}
Then Fronsdal's \cite{F} wave equation is
\begin{equation}
[\Box - (a\partial)(\partial_{a}\partial)+\frac{1}{2}(a\partial)^{2}\Box_{a}]  h^{s}(z;a) = 0
\label{1.3}
\end{equation}

Potentials $h^{(s)}$ admit gauge transformations that leave the wave equations invariant. The gauge functions $\xi^{(s-1)}(z; a)$ are assumed to be traceless
\begin{equation}
\Box_{a}\xi^{(s-1)}(z; a) = 0
\label{1.4}
\end{equation}
and the (classical) gauge transformation is
\begin{equation}
h^{(s)}(z; a) \longrightarrow h^{(s)}(z; a) + (a\partial) \xi^{(s-1)}(z; a)
\label{1.5}
\end {equation}
In the sequel we will not refer to any wave equation.

In \cite{Ber1} the curvature in flat space was defined by
\begin{equation}
R^{(s)}_{\mu_{1}\nu_{1},\mu_{2}\nu_{2}, ... \mu_{s} \nu_{s}}(z) = \\
\prod _{i=1}^{s}(g_{\mu_i}^{\alpha_i}g_{\nu_i}^{\beta_i}
- g_{\mu_i}^{\beta_i}g_{\nu_i}^{\alpha_i}) \quad \\
\partial_{\alpha_1}\partial_{\alpha_2}... \partial_{\alpha_s}
h^{(s)}_{\beta_1\beta_2 ...  \beta_s}(z)
\label{1.6}
\end{equation}
which generalizes corresponding expressions for the electromagnetic case $A = h^{(1)}$ of spin one and the gravitational case $g = h^{(2)}$ of spin two to arbitrary spin.
It is obvious that $R$ is antisymmetric under the exchange inside a single pair
\begin{equation}
(\mu_i, \nu_i) \longrightarrow (\nu_i, \mu_i)
\label{1.7}
\end{equation}
and symmetric under the exchange of two such pairs ($i$ different from $j$)
\begin{equation}
(\mu_i, \nu_i) \longleftrightarrow (\mu_j,\nu_j)
\label{1.8}
\end{equation}
Moreover the curvature $R$ is cyclic in the sense
\begin{equation}
R_{\mu_1\nu_1,\mu_2\nu_2,\mu_3...  \nu_s} + R_{\nu_1\mu_2,\mu_1\nu_2,\mu_3
... \nu_s} + R_{\mu_2\mu_1,\nu_1\nu_2,\mu_3...\nu_s} = 0
\label{1.9}
\end{equation}
Further relations follow from the wave equations which we are not interested in this context.

The de Wit-Freedman curvature $\Gamma^{(s)}$ of $h^{(s)}$ can be derived from
$R^{(s)}$ (1.6) by \cite{DF}
\begin{equation}
\Gamma^{(s)}_{\mu_1\mu_2,...\mu_s; \nu_1\nu_2,...\nu_s} =
\frac{1}{s!} \sum_{P(\nu)}
R^{(s)}_{\mu_1 P(\nu_1),\mu_2 P(\nu_2),...P(\nu_s)} (z)
\label{1.10}
\end{equation}
where the summation is over all permutations $P(\nu)$ of the labels $\nu$.
Thus $\Gamma$ is totally symmetric in each of the $s$-tupels of labels. From now
on we use the harmonic polynomial representation
\begin{equation}
\Gamma^{(s)}(z;a,b) = \prod^{s}_{i=1}(a^{\mu_i}b^{\nu_i})\Gamma^{(s)}
_{\mu_1\mu_2...\mu_s; \nu_1\nu_2...\nu_s} \\
= \prod^{(s)}_{i=1}(a^{\mu_i}b^{\nu_i}) R^{(s)}_{\mu_1 \nu_1,\mu_2 \nu_2,...
\nu_s}(z)
\label{1.11}
\end{equation}
This relation can in fact be inverted so that $R$ and $\Gamma$
contain the same information. This inversion formula has been written
down first in \cite{Ber1} without proof (and with a wrong normalization factor).
Therefore we will prove it here and derive the normalization factor in the Appendix.

Consider the Young diagram with two rows, each of length $s$. The symmetrizer
of either row is denoted $S$ and the antisymmetrizer of the $s$
columns is denoted $A$. Then $AS$ is the symmetrizer of the diagram. It is
idempotent,
\begin{equation}
(AS)^{2} = c_{s}AS
\label{1.12}
\end{equation}
where we leave the normalization $c_{s}$ free. Denoting the $s$-fold derivative of $h^{(s)}$ in (1.6) by $\chi^{(s)}$, we have
\begin{equation}
R^{(s)} = A \chi^{(s)} = AS \chi^{(s)}
\label{1.13}
\end{equation}
On the other hand $\Gamma$ was defined by
\begin{equation}
\Gamma^{(s)} = S R^{(s)}
\label{1.14}
\end{equation}
It follows that
\begin{equation}
A\Gamma^{(s)} = AS R^{(s)} = c_{s} R^{(s)}
\label{1.15}
\end{equation}
In the Appendix it is shown that with our normalizations of $S$ and $A$ the constant is
\begin{equation}
c_{s} = 1
\label{1.16}
\end{equation}
so that the desired inversion formula is
\begin{equation}
R^{(s)} = A \Gamma^{(s)}
\label{1.17}
\end{equation}
or
\begin{equation}
R^{(s)}_{\mu_1\nu_1,...  \mu_s\nu_s}(z) = \\
\prod^{s}_{i=1}(g^{\alpha_i}_{\mu_i}g^{\beta_i}_{\nu_i} - g^{\alpha_i}_{\nu_i}
g^{\beta_i}_{\mu_i}) \Gamma^{(s)}_{\alpha_1\alpha_2...\alpha_s;\beta_1\beta_2...\beta_s}(z)
\label{1.18}
\end{equation}
This completes the proof.

Now we aim at a reformulation of the deWit - Freedman curvature in homogeneous
polynomial form (1.11)
\begin{eqnarray}
\Gamma^{(s)}(z; a,b) &=& \prod_{i=1}^{s} (a^{\alpha_i}b^{\beta_i} - b^{\alpha_i} a^{\beta_i})\partial_{\alpha_1}\partial_{\alpha_2}...\partial_{\alpha_s}
h^{(s)}_{\beta_1\beta_2...\beta_s}(z) \nonumber \\ &=&
\sum_{l=0}^{s} (-1)^{s-l} {s \choose l} (a\partial)^{l} (b\partial)^{s-l}
h^{(s)}_{s-l,l}(z; a,b)
\label{1.19}
\end{eqnarray}
where $s-l,l$ are the degrees of homogeneity in $a$ respectively $b$.

This can be simplified using
\begin{equation}
h^{(s)}_{s-l,l}(z; a,b) = n_{l}^{-1}(b\partial_{a})^{l}
h^{(s)}(z; a)
\label{1.20}
\end{equation}
where
\begin{equation}
n_{l} =s(s-1)(s-2)... (s-l+1)
\label{1.21}
\end{equation}
so that
\begin{equation}
\Gamma^{(s)}(z; b,a) = \sum_{l=0}^{s}\frac{(-1)^{l}}{l!}(a\partial)^{l}
(b\partial)^{s-l}(b\partial_{a})^{l} h^{(s)}(z; a)
\label{1.22}
\end{equation}
We use this expansion to prove gauge invariance of $\Gamma^{(s)}$. This proof is needed in the sequel. Proving gauge invariance of $R^{(s)}$ is trivial.

In fact the gauge variation is
\begin{equation}
\delta\Gamma^{(s)}(z;b,a) = \sum_{l=0}^{s} \frac{(-1)^{l}}{l!} (a\partial)^{l}
(b\partial)^{s-l}(b\partial_{a})^{l}(a\partial) \xi^{(s-1)}(z; a)
\label{1.23}
\end{equation}
where
\begin{equation}
(b\partial_a)^{l}(a\partial)\xi^{(s-1)}(z;a) = (a\partial)(b\partial_a)^{l}\xi^{(s-1)}(z;a) +l(b\partial)(b\partial_a)^{l-1}\xi^{(s-1)}(z;a)
\label{1.24}
\end{equation}
Inserting this into (1.23) and substituting $l$ by $l+1$ in the second term gives zero.

Thus we generalize our ansatz for a curvature on $AdS_{d+1}$ as
\begin{equation}
\Gamma^{(s)}(z;b,a) = \sum_{k=0}^\infty L^{-2k} \Gamma^{(s)}_{k}
(z;b,a)
\label{1.25}
\end{equation}
including finitely many $1/L^{2}$ correction terms only and the leading term at $k=0$
\begin{equation}
\Gamma^{(s)}_{0}(z;b,a) = \sum_{l=0}^{s} \frac{(-1)^{l}}{l!}
(a\nabla)^{l}(b\nabla)^{s-l}(b\partial_a)^{l} h^{(s)}(z;a)
\label{1.26}
\end{equation}
which is obtained from (1.22) by replacing space derivatives by covariant derivatives and fixing their order
(another order could also do it). In an analogous fashion and following the seminal article \cite{F}
we make an ansatz for generalized Christoffel symbols by a polynomial expression in $L^{-2}$ and for all $1\leq m\leq
s-1$
\begin{equation}
\Gamma^{(m,s)}(z;b,a) = \sum_{k\geq 0} L^{-2k} \Gamma_{k}^{(m,s)}(z;b,a)
\label{1.27}
\end{equation}
\begin{equation}
\Gamma^{(m,s)}_{0}(z;b,a) = \sum_{l=0}^{m} \frac{(-1)^{l}}{l!} (a\nabla)^{l}(b\nabla)^{m-l}(b\partial_{a})^{l}
h^{(s)}(z;a)
\label{1.28}
\end{equation}
\begin{equation}
\Gamma^{(s,s)}(z;b,a) = \Gamma ^{(s)}(z;b,a)
\label{1.29}
\end{equation}
We use as ansatz the following representation for $\Gamma_{k}^{(m,s)}$ including the case $m=s$
\begin{eqnarray}
\Gamma_{k}^{(m,s)} &=& \sum_{r_1,r_2,r_3} \sum_{l=2r_1+r_2}^{m-r_2-2r_3}\frac{(-1)^{l}}{l!} A_{r_1r_2r_3}^{(l)} (a^{2})^{r_1}(ab)^{r_2}(b^{2})^{r_3} \nonumber\\
&&(a\nabla)^{l-2r_1-r_2} (b\nabla)^{m-l-r_2-2r_3}(b\partial_{a})^{l} h^{(s)}(z; a) \nonumber\\
&& (r_{i} \geq 0, r_1+r_2+r_3 = k)
\label{1.30}
\end{eqnarray}
In the case of the curvature $\Gamma^{(s)}$ the requirement of gauge invariance
\begin{equation}
\delta \Gamma^{(s)}(z;b,a) = 0
\label{1.30}
\end{equation}
the main motivation being that in $AdS$ length scale $L,a,b$ have scale $+1$ whereas $\nabla$ has scale $-1$.
So a power $L^{-2k}$ in the denominator necessitates an invariant compensator in the numerator. The covariant derivative cannot be used for this purpose.

The coefficients $A$ in (1.30) must be determined by requiring gauge invariance. In the case of the curvature
$\Gamma^{(s)}$ the requirement of gauge invariance
\begin{equation}
\delta\Gamma^{(s)}(z;b,a) = 0
\label{1.31}
\end{equation}
and in the case of the Christoffel symbols (i.e. for $m<s$) tracelessness with respect to $b$ of the gauge variation
\begin{equation}
Tr_{b}\delta\Gamma^{(m,s)}(z;b,a) = 0
\label{1.32}
\end{equation}
are postulated. In the flat case and for $m=2$ the second order Fronsdal equation of motion for a spin $s$ gauge field results from (see [3, 2])
\begin{equation}
Tr_{b} \Gamma^{(2,s)}(z;b,a) = 0
\label{1.33}
\end{equation}
which is gauge invariant (only in the flat case) due to
\begin{eqnarray}
\delta\Gamma^{(m,s)} (z;b,a) = \frac{(-1)^{m}}{m!}(a\partial)^{m+1}(b\partial_{a})^{m}\xi^{(s-1)}\\
\label{1.34}
Tr_{b}\delta\Gamma^{(m,s)}(z;b,a) = \frac{(-1)^{m}}{(m-2)!}(a\partial)^{m+1}(b\partial)^{m-2} Tr_{a}\xi^{(s-1)}(z;a)
= 0 \nonumber\\
\label{1.35}
\end{eqnarray}

These requirements can be transformed into recursive equations of $\Gamma^{(m,s)}_{k}, \Gamma^{(s)}_{k}$. The derivation and solution of these is the topics of this article. Though the task seems different in both cases, we can treat it in parallel a long time. We will display the mathematical apparatus for an elegant treatment of these problems first.

\setcounter{equation}{0}
\section{Differential algebra on symmetric tensors}
In this section we develop the Lie algebraic algorithm of covariant derivatives acting on symmetric
higher spin fields on anti-deSitter spaces. Application of gradients to these fields produce tensor fields which are symmetric in two different sets of labels, we call them "bisymmetric" tensors $t_{rs}(z;a,b)$. These are homogeneous polynomials in $a$ of degree $r$ and in $b$ of degree $s$. The
vectors $a$, $b$ belong to the tangential space at $z$, $TAdS(z)$. The linear
space of such $C_{\infty}$ fields over $AdS$ is denoted $T_{rs}(a,b)$ and their union is
\begin{equation}
\bigoplus_{r,s}T_{rs}(a,b) = T(a,b)
\label{2.1}
\end{equation}
On such space $T(a,b)$ we can act with the differential operator algebra
which in the sequel will be our main tool.

The first class consists of $a$-gradients, $b$-gradients, $a$-divergences,
$b$-divergences etc., e.g.
\begin{eqnarray}
a\textnormal{-gradient}: (a\nabla)t_{rs}(z;a,b) \in T_{r+1,s}(a,b)
\label{2.2} \\
a\textnormal{-divergence}: (\nabla\partial_a)t_{rs}(z;a,b) \in T_{r-1,s}(a,b)
\label{2.3}
\end{eqnarray}
Another class of operators consists of purely tensorial operators such as
\begin{equation}
(a\partial_b)t_{rs}(z;a,b) \in T_{r+1,s-1}(a,b)
\label{2.4}
\end{equation}
or the Euler operators $(a\partial_a)$ and $(b\partial_b)$.

On curved spaces the commutator of two gradients
\begin{equation}
\alpha_1 = [(a\nabla),(b\nabla)] = -\beta_1
\label{2.5}
\end{equation}
is of great importance and, for $AdS$ spaces of radius $L$, assumes the
simple form\footnote {This definition of the commutator is in agreement with the following conventions for the  Euclidian $AdS_{d+1}$ metric and curvature\begin{eqnarray}
&&ds^{2}=g_{\mu \nu }(z)dz^{\mu }dz^{\nu
}=\frac{L^{2}}{(z^{0})^{2}}\delta _{\mu \nu }dz^{\mu }dz^{\nu
},\quad \sqrt{g}=\frac{L^{d+1}}{(z^{0})^{d+1}}\;,
\notag  \label{ads} \\
&&\left[ \nabla _{\mu },\,\nabla _{\nu }\right] V_{\lambda }^{\rho }=R_{\mu
\nu \lambda }^{\quad \,\,\sigma }V_{\sigma }^{\rho }-R_{\mu \nu \sigma
}^{\quad \,\,\rho }V_{\lambda }^{\sigma }\;,  \notag \\
&&R_{\mu \nu \lambda }^{\quad \,\,\rho
}=-\frac{1}{(z^{0})^{2}}\left( \delta _{\mu \lambda }\delta _{\nu
}^{\rho }-\delta _{\nu \lambda }\delta _{\mu }^{\rho }\right)
=-\frac{1}{L^{2}}\left( g_{\mu \lambda }(z)\delta _{\nu
}^{\rho }-g_{\nu \lambda }(z)\delta _{\mu }^{\rho }\right) \;,  \notag \\
&&R_{\mu \nu }=-\frac{d}{(z^{0})^{2}}\delta _{\mu \nu }=-\frac{d}{L^{2}}%
g_{\mu \nu }(z)\quad ,\quad R=-\frac{d(d+1)}{L^{2}}\;.  \notag
\end{eqnarray} }
\begin{equation}
\alpha_1 = \frac{1}{L^{2}} \{b^{2}(a\partial_b)-a^{2}(b\partial_a)
-(ab)[(b\partial_b)- (a\partial_a)]\}
\label{2.6}
\end{equation}
In the same fashion we define
\begin{eqnarray}
\alpha_2 = [(a\nabla),\alpha_1] = \frac{1}{L^{2}}\{a^{2}(b\nabla) -(ab)(a\nabla)
\}\\
\label{2.7}
\beta_2 = [(b\nabla), \beta_1] = \frac{1}{L^{2}}\{b^{2}(a\nabla) -(ab)(b\nabla)
\}
\label{2.8}
\end{eqnarray}
so that by replacing $a$ by $b$ and vice versa,  $\alpha_{k}$ and $\beta_{k}$
exchange their roles.
Define higher order commutators by
\begin{eqnarray}
[(a\nabla), \alpha_k] = \alpha_{k+1} \\
\label{2.9}
[(b\nabla), \beta_k] = \beta_{k+1}
\label{2.10}
\end{eqnarray}
then we can easily see that a repetitive structure arises
\begin{eqnarray}
\alpha_{2n+1} = \left(\frac{a^{2}}{L^{2}}\right)^{n}\alpha_1\\
\label{2.11}
\alpha_{2n} = \left(\frac{a^{2}}{L^{2}}\right)^{n-1}\alpha_2 \\
\label{2.12}
\beta_{2n+1} = \left(\frac{b^{2}}{L^{2}}\right)^{n}\beta_1\\
\label{2.13}
\beta_{2n} = \left(\frac{b^{2}}{L^{2}}\right)^{n-1}\beta_2
\label{2.14}
\end{eqnarray}
Next we expand
\begin{eqnarray}
[(b\nabla)^{n}, (a\nabla)] = \sum_{k=1}^{n} C^{n}_{n-k}(b\nabla)^{n-k}\beta_k \\
\label{2.15}
[(b\nabla)^{n}, \beta_1] = \sum_{k=1}^{n} D_{n-k}^{n} (b\nabla)^{n-k}\beta_
{k+1} \\
\label{2.16}
[(b\nabla)^{n}, \beta_2] = \sum_{k=1}^{n} E_{n-k}^{n} (b\nabla)^{n-k} \beta_{k+2}
\label{2.17}
\end{eqnarray}
The summation starts with $k=1$ in each case because by commutation the power of $(b\nabla)^{n}$ is lowered at least by one.
By Jacobi's identity we obtain immediately for $k>0$
\begin{equation}
C_{n-k}^{n} = D_{n-k}^{n} = E_{n-k}^{n}
\label{2.18}
\end{equation}
A recursion relation is obtained from
\begin{equation}
[(b\nabla)^{n+1},(a\nabla)] = (b\nabla)[(b\nabla)^{n}, (a\nabla)]
+ [(b\nabla),(a\nabla)](b\nabla)^{n}
\label{2.19}
\end{equation}
which is solved by
\begin{equation}
C_{n-k}^{n} = [{n \choose k} -\delta_{k,0}](-1)^{k-1}
\label{2.20}
\end{equation}
It follows finally that
\begin{eqnarray}
[(b\nabla)^{n}, (a\nabla)] = \left\{ \sum_{k=1}^{[\frac{n+1}{2}]}{ n \choose 2k-1}(b\nabla)^{n-2k+1}
\left(\frac{b^{2}}{L^{2}}\right)^{k-1}\right\} \beta_1 \nonumber \\
- \left\{\sum_{k=1}^{[\frac{n}{2}]}{ n \choose 2k }(b\nabla)^{n-2k} \left(\frac{b^{2}}{L^{2}}\right)^{k-1}\right\}\beta_2
\label{2.21}
\end{eqnarray}
This formula is the basis for the recursion relations for $\Gamma^{(s)}_{k}$.

\setcounter{equation}{0}
\section{The gauge variation of $\Gamma^{(m,s)}$}
\subsection{The gauge variation of $\Gamma^{(m,s)}_0$}

In order to derive equations from the gauge invariance postulate, we have to make
sufficiently detailed studies of the gauge variations of $\Gamma_{0}^{(m,s)}$ and $\Gamma_{k}^{(m,s)},
k\geq 1$. The first object gives the inhomogeneous terms in the linear system of equations, the second objects
give the matrix (linear operator) acting on the unknowwn coefficients $A$. In the sequel we will call a term "level $k$", if its differential operators are "canonically" ordered according to
\begin{equation}
(a\nabla)^{r}(b\nabla)^{s}(b\partial_{a})^{t}
\label{3.1}
\end{equation}
and its coefficient is $O(L^{-2k})$. The levels define an ordering scheme
for the curvature and its gauge variation. We assume $m$ arbitrary $\leq s$.

We start from $\Gamma^{(m,s)}_{0}$ (1.28) and decompose its gauge variation
into contributions of the different levels. From (1.28) we have
\begin{equation}
\delta\Gamma^{(m,s)}_{0} = \sum_{l=0}^{m} \frac{(-1)^{l}}{l!} (a\nabla)^{l}
(b\nabla)^{m-l}(b\partial_{a})^{l}(a\nabla)\xi^{(s-1)}(z;a)
\label{3.2}
\end{equation}
which can, following the treatment of the flat case, be brought
into the form
\begin{equation}
\delta\Gamma^{(m,s)}_{0} = \sum_{l=0}^{m} \frac{(-1)^{l}}{l!} (a\nabla)^{l}
[(b\nabla)^{m-l}, (a\nabla)](b\partial_{a})^{l}\xi^{(s-1)}(z;a) +\frac{(-1)^{m}}{m!}(a\nabla)^{m+1}(b\partial_{a})^{m}
\xi^{(s-1)}(z;a)
\label{3.3}
\end{equation}
The last term vanishes for $m=s$ or after taking a $b$-trace. We can neglect it for both the curvature and the Christoffel symbols.

Using (2.21) the commutator can be evaluated as
\begin {eqnarray}
\delta\Gamma^{(m,s)}_{0} &=& \sum_{k > 0}\sum_{l=0}^{m} \frac{(-1)^{l}}{l!}
(a\nabla)^{l} \left(\frac{b^{2}}{L^{2}}\right)^{k-1} \left[{m-l \choose 2k-1 }(b\nabla)^{m-l-2k+1}
 \beta_{1} \right.\nonumber \\ &-& \left.  {m-l \choose 2k} (b\nabla)^{m-l-2k}
 \beta_{2}\right]  (b\partial_{a})^{l}\xi^{(s-1)}
\label{3.4}
\end{eqnarray}
From (2.8) we recognize that the $\beta_{2}$ term is not canonically
ordered. Performing the ordering we obtain
\begin{equation}
(b\nabla)^{m-l-2k}\beta_{2} = L^{-2}[b^{2}(a\nabla)(b\nabla)^{m-l-2k}
- (ab)(b\nabla)^{m-l-2k+1}] + \frac{b^{2}}{L^{2}}[(b\nabla)^{m-l-2k},(a\nabla)]
\label{3.5}
\end{equation}
The first term ($\beta_{1}$ term) in the square bracket of (3.4) contributes only to the level $k$ but the second term ($\beta_{2}$ term) gives rise to infinitely many terms of level larger than $k$.

Consider now the $\beta_{1}$ term of (3.4). From (2.6) we obtain
\begin{equation}
\beta_{1}(b\partial_{a})^{l}\xi^{(s-1)} = L^{-2}
[a^{2}(b\partial_{a})^{l+1}+(2l-s+1)(ab)(b\partial_{a})^{l} - l(s-l)
 b^{2}(b\partial_{a})^{l-1}] \xi^{(s-1)}
\label{3.6}
\end{equation}
We cast the terms of level $k$ resulting into the general form
\begin{equation}
L^{-2k}\sum_{l=0}^{m}\sum_{r_1 r_2 r_3} \frac{(-1)^{l}}{l!} Q^{(k,l)}_{r_1 r_2 r_3}(a^{2})^{r_{1}}(ab)^{r_{2}}
(b^{2})^{r_{3}} (a\nabla)^{l-2r_{1}-r_{2}+1} (b\nabla)^{m-l-r_{2}-2r_{3}}
(b\partial_{a})^{l} \xi^{(s-1)}
\label{3.7}
\end{equation}
and the $r_{i}$ are submitted to
\begin{equation}
\sum_{i=1}^{3} r_{i} = k
\label{3.8}
\end{equation}
The only nonvanishing coefficients $Q$ follow from (3.5), (3.6)
\begin{eqnarray}
Q_{00,k}^{(k,l)} &=&  {m-l-1 \choose 2k-1}(s-l-1)
- {m-l \choose 2k}\\
\label{3.9}
Q_{01,k-1}^{(k,l)} &=&   {m-l\choose 2k-1} (2l-s+1) +
{m-l \choose 2k} \\
\label{3.10}
Q_{10,k-1}^{(k,l)} &=& - {m-l+1 \choose 2k-1 } l
\label{3.11}
\end{eqnarray}
In the sequel we will replace $k$ by $k_1$, eventually.

Now we consider the commutator term in (3.5) applying (2.21) again.
Introducing a new label $k_{2}$ we obtain the expansion
\begin{eqnarray}
\sum_{k_1,k_2\geq 1}\sum_{l=0}^{m}\frac{(-1)^{l+1}}{l!}&&(a \nabla)^{l}
{m-l \choose 2k_1}\left(\frac{b^{2}}{L^{2}}\right)^{k_1 + k_2 -1}
\left\{{m-l-2k_1 \choose 2k_2 -1}(b\nabla)^{m-l-2k_1 -2k_2 +1}\beta_{1}\right.\nonumber\\ &&
\left.- {m-l-2k_1 \choose 2k_2 }(b\nabla)^{m-l-2k_1-2k_2} \beta_{2}\right\}(b\partial_{a})^{l}\xi^{(s-1)}
\label{3.12}
\end{eqnarray}
On the level $k_1+k_2$ remains
\begin{equation}
L^{-2(k_1+k_2)}\sum_{l}\frac{(-1)^{l}}{l!} \sum_{r_{i}} Q_{r_1 r_2 r_3}^{(k_1,k_2,l)} (a^{2})^{r_1} (ab)^{r_2} (b^{2})^{r_3} (a\nabla)^{l-2r_1 -r_2+1}(b\nabla)^{m-l-r_2- 2r_3 }(b\partial_{a})
^{l} \xi^{(s-1)}
\label{3.13}
\end{equation}
where the only nonvanishing coefficients are
\begin{equation}
Q_{00,k_1+k_2}^{(k_1,k_2,l)}  =  - {m-l-1 \choose 2k_1 }{m-l-2k_1-1 \choose 2k_2-1} (s-l-1) +
 {m-l \choose 2k_1 }{m-l-2k_1 \choose 2k_2 }
\label{3.14}
\end{equation}
and
\begin{equation}
Q_{01,k_1+k_2-1}^{(k_1,k_2,l)}  =  - {m-l \choose 2k_1}{m-l-2k_1 \choose 2k_2 -1} (2l-s +1) -
{m-l \choose 2k_1}{m-l-2k_1 \choose 2k_2}
\label{3.15}
\end{equation}
and finally
\begin{equation}
Q_{10,k_1+k_2-1}^{(k_1,k_2,l)} =  + { m-l+1 \choose 2k_1}{m-l-2k_1 +1 \choose 2k_2 -1} l
\label{3.16}
\end{equation}
We notice that all products of two binomial coefficients
can be expressed by multinomial coefficients.

Continuing this procedure we obtain the contribution at level $K$ by
summing over all partitions $\{k_1,k_2,k_3...k_{n}\}$ for which
\begin{equation}
k_{i} \geq 1, \sum_{i=1} ^{n} k_{i} = K
\label{3.17}
\end{equation}
For all such partitions the contribution is then
\begin{eqnarray}
\left(\frac{b}{L}\right)^{2K-2} \sum_{k_{i}\geq 1, \sum k_{i} =K} \sum_{l=0}^{m} \frac{(-1)^{l+n-1}}{l!}
(a\nabla)^{l} \left[{m-l \choose 2k_1,2k_2,...2k_{n}-1, m-l-2K +1} \right.\nonumber \\ (b\nabla)^{m-l-2K+1}\beta_1
\left. - {m-l  \choose 2k_1,2k_2,... 2k_{n}, m-l-2K} (b\nabla)^{m-l-2K} \beta_2 \right]
(b\partial_{a})^{l} \xi^{(s-1)}
\label{3.18}
\end{eqnarray}
where we used multinomial coefficients. Neglecting the commutator in (3.18)
we obtain in terms of coefficients $Q_{r_1 r_2 r_3}^{(k_1, k_2,...k_{n},l)}$
for a fixed partition
\begin{equation}
L^{-2K}\sum_{l=0}^{m}\frac{(-1)^{l}}{l!} Q_{(r_1 r_2 r_3)}^{(k_1,k_2,... k_{n},l)}
(a^{2})^{r_1} (ab)^{r_2} (b^{2})^{r_3} (a\nabla)^{l-2r_1 -r_2 +1}
(b\nabla)^{m-l-r_2 -2r_3} (b\partial_{a})^{l} \xi^{(s-1)}
\label{3.19}
\end{equation}
where the only nonvanishing coefficients are
for fixed $l$ and fixed partition
\begin{eqnarray}
Q_{00,K}^{(k_1,k_2,...k_{n},l)} &=&(-1)^{n}\left[ - {m-l-1 \choose 2k_1, 2k_2,...
2k_{n-1},2k_{n} -1, m-l -2K} (s-l-1)\right. \nonumber \\
 &+&\left.{m-l \choose 2k_1, 2k_2,... 2k_{n}, m-l-2K}\right] \\
\label{3.20}
Q_{01,K-1}^{(k_1,k_2,...k_{n}, l)} &=& (-1)^{n+1} \left[ {m-l\choose 2k_1,2k_2,...
2k_{n-1},2k_{n} -1, m-l -2K+1}(2l-s+1) \right.\nonumber \\  &+& \left.{m-l \choose 2k_1, 2k_2, ... 2k_{n}, m-l-2K }\right]  \\
\label{3.21}
Q_{10,K-1}^{(k_1,k_2,...k_{n},l)}& =& (-1)^{n} {m-l+1 \choose 2k_1,2k_2,...
2k_{n-1},2k_{n}-1, m-l-2K+2} l
\label{3.22}
\end{eqnarray}

\subsection{The gauge variation of $\Gamma_{k}^{(m,s)}$ for $k>0$}
That part of $\Gamma^{(m,s)}$ which is of level $k$ is introduced by the ansatz (see (1.30))
\begin{equation}
L^{2k}\Gamma_{k}^{(m,s)} = \sum_{l=0}^{m} \frac{(-1)^{l}}{l!} \sum_{r_{i}}
A_{r_1 r_2 r_3} ^{(l)}(a^{2})^{r_1}(ab)^{r_2} (b^{2})^{r_3} (a\nabla)^{l-2r_1 -r_2} (b\nabla)^{m-l-r_2 - 2r_3}(b\partial_{a})^{l} h^{(s)}
\label{3.23}
\end{equation}
where the sum over the $r_{i}$ is restricted to $\sum r_{i} = k$.
The strategy is to derive the unknown coefficients $A$ from the known coefficients $Q$ by requiring (1.31),(1.32).
The idea is to postulate (1.31) for both cases, the curvature and the Christoffel symbols, tentatively, in order to derive a system of difference equations for the $A$. Then one solves these if they are solvable. Otherwise one determines the obstruction of solvability. Only for $m<s$ the obstructions are effective as we shall see (after an intricate analysis) and the weaker constraint (1.32) is applied

The sum over $l$ for given ${r_{i}}$ is restricted by
\begin{equation}
2r_1 +r_2 \leq l \leq m-r_2 - 2r_3
\label{3.24}
\end{equation}
which implies
\begin{equation}
k \leq [s/2]
\label{3.25}
\end{equation}
The gauge variation is easily obtained in the form
\begin{eqnarray}
&&L^{2k}\delta \Gamma_{k}^{(m,s)} = \sum_{l=0}^{m} \sum_{r_{i}} \frac{(-1)^{l}}{l!}
(a^{2})^{r_1}(ab)^{r_2} (b^{2})^{r_3}
\nonumber\\ &&\Big\{(A_{r_1 r_2 r_3}^{(l)} - A_{r_1 r_2 r_3}^{(l+1)}) (a\nabla)^{l-2r_1 -r_2+1} (b\nabla)^{m-l-r_2 -2r_3}
\nonumber\\ &&+A_{r_1 r_2 r_3}^{(l)} (a\nabla)^{l-2r_1-r_2} [(b\nabla)^{m-l-r_2 -2r_3},
(a\nabla)]\Big\} (b\partial_{a})^{l} \xi^{(s-1)}\nonumber\\
\label{3.26}
\end{eqnarray}
Thus $\delta \Gamma_{k}^{(m,s)}$ decomposes into a level $k$ term and terms
of level $K =k+\kappa, \kappa >0$ by expansion of the commutator into contributions of different ordered partitions $\{k_2, k_3... k_{n}\}$ of $\kappa$
\begin{equation}
\kappa = k_2 +k_3 +... k_{n}, k_{i} >0
\label{3.27}
\end{equation}
By the now standard analysis we get for the first commutator and the fixed partition $\{k_{2}\}$
\begin{eqnarray}
L^{-2k}\left(\frac{b}{L}\right)^{2k_{2} -2}\sum_{l=0}^{m} \frac{(-1)^{l}}
{l!}
\sum_{r_{i}} A_{r_1 r_2 r_3}^{(l)}(a^{2})^{r_1}(ab)^{r_2}(b^{2})^{r_3}
(a\nabla)^{l-2r_1 -r_2}
\nonumber \\
\left\{{ m-l-r_2 -2r_3 \choose 2k_2-1} (b\nabla)^{m-l-r_2-2r_3-2k_{2} +1}\beta_1\right.\nonumber\\
\left.-{ m-l-r_2 -2r_3 \choose  2k_2 }(b\nabla)^{m-l-r_2 -2r_3-2k_2}\beta_2 \right\} (b\partial_{a})^{l}\xi^{(s-1)} \label{3.28}
\end{eqnarray}
After neglecting the new commutator generated by $\beta_2$ and inserting (3.5), (3.6) and simplifying we get for this partition $\{k_{2}\}$ the expression with the $A$s of level $k$ added to an equation for the level $k+k_2$
\begin{eqnarray}
\sum_{l=0}^{m}\frac{(-1)^{l}}{l!} L^{-2K}\sum_{r_{i}}(a^{2})^{r_1} (ab)^{r_{2}}(b^{2})^{r_3 +
k_{2} -1} (a\nabla)^{l-2r_1-r_2+1} (b\nabla)^{m-l-r_2- 2r_3-2k_{2}+2}\nonumber \\
\left\{\left[ A_{r_1, r_2, r_3 -1}^{(l+1)} {m-l-r_2-2r_3+1 \choose 2k_2 -1}(s-l-1)-  A_{r_1,r_2,r_3-1}^{(l)}{m-l-r_2-2r_3 +2 \choose 2k_2}\right]\right. \nonumber\\
+A_{r_1,r_2-1,r_3}^{(l)}\left[{m-l-r_2-2r_3+1 \choose 2k_2-1}(2l-s+1) +{m-l-r_2-2r_3+1 \choose 2k_2}\right] \nonumber\\
\left.- A_{r_1 -1,r_2, r_3}^{(l-1)} {m-l-r_2-2r_3+1 \choose 2k_2 -1} l \right\}(b\partial_{a})^{l} \xi^{(s-1)}\qquad\qquad
\label{3.29}
\end{eqnarray}
where the $r_{i}$ are restricted to
\begin{equation}
\sum r_{i} = k +1
\label{3.30}
\end{equation}
The recursion relations for the coefficients $A$ at level one and two are determined and solved in the subsequent section. For general partitions $\{k_2,k_3...k_{n}\}$ the relevant formulae are easily obtained but too clumsy to be given here.

\setcounter{equation}{0}
\section{Setting up and solving the recursion relations}
\subsection{$\Gamma_{1}^{(m,s)}$ and the concept of obstruction}

It is natural to solve the condition of gauge invariance level by level
in increasing order. The equations we have to solve have the following structure
\begin{eqnarray}
 A_{r_1r_2r_3}^{(l+1)} - A_{r_1r_2r_3}^{(l)}  &=&  Q_{r_1r_2r_3}^{(k,l)} + \sum_{k_1 =1}^{k-1} Q_{r_1r_2r_3}^{(k_1,k-k_1,l)} +\sum_{k_1,k_2 \geq 1,k_1+k_2\leq k-1} Q_{r_1r_2r_3}^{(k_1,k_2,k-k_1-k_2,l)}+...\nonumber\\
&&+ \sum_{l',\sigma_1,\sigma_2,\sigma_3} q_{\sigma_1\sigma_2\sigma_3}^{(l,l')}A_{r_1-\sigma_1,r_2-\sigma_2,r_3-\sigma_3}^{(l')}\nonumber\\
&&(0\leq \sigma_{i} \leq r_{i}, 1\leq \sigma_1+\sigma_2+\sigma_3 \leq k-1, r_1+r_2+r_3 = k)
\label{4.1}
\end{eqnarray}
where the r.h.s. $A$-terms exist if $k\geq 2$, result from formulas such as (3.29), and must be summed over all
partitions. In the simplest case $k=1$ they do not arise. In that case we have
\begin{equation}
A_{r_1r_2r_3}^{(l+1)} - A_{r_1r_2r_3}^{(l)} = Q_{r_1r_2r_3}^{(1,l)},\quad
(r_1+r_2+r_3 = 1)
\label{4.2}
\end{equation}
This case $k=1$ is studied first.

In the sequel we use Pascal's summation formula which for
\begin{equation}
A^{(l+1)} - A^{(l)} = {a-l \choose n} =\frac{[a-l]_{n}}{n!}
\label{4.3}
\end{equation}
implies
\begin{equation}
A^{(l)} = - {a+1-l \choose n+1} + \textnormal{constant}
\label{4.4}
\end{equation}
We insert the expressions (3.9)-(3.11) for $k=1$ into (4.1)
\begin{eqnarray}
(a)\qquad Q_{001}^{(1,l)} &=& \frac{1}{2}(m-l-1)(2s-m-l-2) \\
\label{4.5}
(b)\qquad Q_{010}^{(1,l)} &=& \frac{1}{2}(m-l)(3l -2s+m+1) \\
\label{4.6}
(c)\qquad Q_{100}^{(1,l)} &=& - l(m-l+1)
\label{4.7}
\end{eqnarray}
In the case $k=1$ there are no lower level $A$-terms.

These coefficients are nonzero in the intervals
\begin{eqnarray}
&(a)& \qquad 0 \leq l \leq m-2 \\
\label{4.8}
&(b)&\qquad 0 \leq l \leq m-1 \\
\label{4.9}
&(c)&\qquad 1 \leq l \leq m
\label{4.10}
\end{eqnarray}
From the ansatz (1.30),(3.23) which excludes negative powers of $a^{2}, (ab), b^{2}$ and
$a\nabla, b\nabla$ follows (see (3.25))
\begin{equation}
2r_1 + r_2 := l_{\min} \leq l\leq l_{\max}: = m -r_2 -2r_3
\label{4.11}
\end{equation}
Moreover from the multinomials in $Q$ (3.20)-(3.22) and from (4.11) we can deduce the support of $Q^{(k,l)}$ in $l$
\begin{equation}
\max\{0, l_{\min}-1\} \leq l \leq  l_{\max}
\label{4.12}
\end{equation}
which exactly reproduces (4.8)-(4.10).

Boundary conditions of the difference equations are obtained from those equations that have only one $A$ on the left hand side, namely
\begin{eqnarray}
&(a)&\qquad  -A^{(m-2)} = Q^{(1,m-2)}\\
\label{4.13}
&(b)&\qquad  -A^{(m-1)} = Q^{(1,m-1)}, A^{(1)} = Q^{(1,0)}\\
\label{4.14}
&(c)&\qquad  -A^{(m)} = Q^{(1,m)}, A^{(2)} = Q^{(1,1)}
\label{4.15}
\end{eqnarray}
so that we have to deal with two boundary conditions in cases $(b)$ and $(c)$,
which, except under special circumstances, is one too many.
The generalization to arbitrary $k$ is straightforward. Only in the case
\begin{equation}
(r_1,r_2,r_3) = (0, 0, k)
\label{4.16}
\end{equation}
we have one, and in all other cases we have two boundary conditions. But two boundary conditions are in general
incompatible with first order difference equations. As usual we define $A_{r_1r_2r_3}^{(l)}$ to be zero outside
its support (4.11). The incompatibility can then be expressed by an "obstruction" $\Delta_{r_1r_2r_3}$ through
\begin{equation}
\Delta_{r_1r_2r_3} = A_{r_1r_2r_3}^{(l_{\max})} + \sum_{all partitions}Q_{r_1r_2r_3}^{(...,l_{\max})} +\textnormal{ A-terms of lower level as in (4.1)}
\label{4.17}
\end{equation}
which, if the upper boundary condition is satisfied, vanishes identically.

For the cases with two boundary conditions we determine the "semi-solution" satisfying only the lower boundary condition. For $k = 1$ this gives
\begin{eqnarray}
(a)\qquad A_{001}^{(l)} &=& -{m-l\choose 3} -(s-m){m-l\choose 2}\qquad (\textnormal{proper solution})\\
\label{4.18}
(b)\qquad A_{010}^{(l)} &=& (-l+s-m){m-l+1\choose 2} -(s-m){m+1\choose 2}\nonumber\\&&(\textnormal{semi-solution})\\
\label{4.19}
(c)\qquad A_{100}^{(l)} &=& 2{l\choose 3} -m {l\choose 2}\qquad (\textnormal{semi-solution})
\label{4.20}
\end{eqnarray}
and the obstruction is in either case
\begin{eqnarray}
(b)\qquad \Delta_{010} &=& A_{010}^{(m-1)} + Q_{010}^{(1,m-1)} = {m+1 \choose 2}(m-s)\\
\label{4.21}
(c)\qquad \Delta_{100} &=& A_{100}^{(m)} +Q_{100}^{(1,m)} = - {m+2 \choose 3}
\label{4.22}
\end{eqnarray}
These obstruction functions enter into the gauge variation of $\Gamma$ which to the order $O(L^{-2})$
gives
\begin{eqnarray}
  && L^{2}\delta \Gamma^{(m)(s)}_{0}(z;a,b)+\delta \Gamma^{(m)(s)}_{1}(z;a,b)= \nonumber\\
  &&\frac{(-1)^{m}}{m!}(a\nabla)^{m-1}\left[L^{2}(a\nabla)^{2}
  -\frac{1}{6}m(m+1)(m+2)a^{2}\right](b\partial_{a})^{m}\xi^{(s-1)}(z;a)\nonumber\\
  &&+(ab)(s-m)\frac{(-1)^{m}m(m+1)}{2 (m-1)!}(a\nabla)^{m-1}
  (b\partial_{a})^{m-1}\xi^{(s-1)}(z;a)\label{4.23}
\end{eqnarray}
From this result we conclude that for the curvature with  $s=m$ and for the level $k=1$ we have obtained a gauge invariant expression, since
\begin{equation}
(b\partial_{a})^{s}\xi^{(s-1)}(z;a) = 0                                                                                 \label{4.24}
\end{equation}

For the Christoffel symbols we have to make the $b$-trace of the gauge variation traceless
\begin{eqnarray}
  &&\frac{1}{2}\Box_{b}\left\{ L^{2}\delta \Gamma^{(m,s)}_{0}(z;a,b)
  +\delta \Gamma^{(m,s)}_{1}(z;a,b)\right\}= \nonumber\\
  && \frac{(-1)^{m}}{2(m-2)!}m(m+1)(s-m)(s-m+1)
  (a\nabla)^{m-1}(b\partial_{a})^{m-2}\xi^{(s-1)}(z;a) .\label{4.25}
\end{eqnarray}
This trace we can cancel only adding an additional $b^{2}$
term
\begin{equation}\label{4.26}
    {\mathcal L_{1}[h^{(s)}]}=\frac{b^{2}m(m+1)(s-m)(s-m+1)}{d+2m-3}\sum^{m-1}_{l=1}
    \frac{(-1)^{l}}{2(l-1)!}(a\nabla)^{l-1}
    (b\nabla)^{m-l-1}(b\partial_{a})^{l-1}h^{(s)}(z;a)
\end{equation}
The trace of the gauge variation of this term cancels exactly the r.h.s of (\ref{4.25}).

Effectively this term leads to a $d$-dependent shift of $A_{001}^{(l)}$ coefficients independent of $l$,
which amounts to a change in the (single) boundary condition for the difference equation for these terms.
The ansatz (1.30), (3.23) remains unchanged. In (\ref{3.29}) we must, however, insert now the modified $A_{00k}^{(l)}*$
\begin{equation}
A_{001}^{(l)} \Rightarrow A_{001}^{(l)}* = A_{001}^{(l)} - {m+1\choose 2}\frac{(s-m)(s-m+1)}{d+2m-3}
\label{4.27}
\end{equation}
At the end we change the notation and write again $A_{001}^{(l)}$ for $A_{001}^{(l)}*$.

The case $m=2$ is special because in this case
all our sums degenerate and we have only the boundary conditions for the remaining
$A_{001}^{(0)}, A_{010}^{(1)}, A_{100}^{(2)}$ coefficients. Instead of (\ref{4.23}) we have here (with $B = -A^{(1)}_{010}$)
\begin{eqnarray}
  && \delta \left\{L^{2}\Gamma^{(2,s)}_{0}(z;a,b)+\left[(2-s)b^{2}
  +B(ab)(b\partial_{a})-a^{2}(b\partial_{a})^{2}\right]h^{(s)}(z;a)\right\}= \nonumber\\
  &&(ab)\left[(B+3-2s)(b\nabla)+(B-3+s)
  (a\nabla)(b\partial_{a})\right]\xi^{(s-1)}(z;a)
\label{4.28}
\end{eqnarray}
which can be made traceless if we choose
\begin{equation}\label{4.29}
    B = -\frac{1}{s}(s^{2}-6s+6)
\end{equation}
So the correct second Christoffel symbol is
\begin{eqnarray}
  \Gamma^{(2,s)}(z;a,b)&=&(b\nabla)^{2}h^{(s)}(z;a)-(a\nabla)\left[(b\nabla)(b\partial_{a})
  -\frac{1}{2}(a\nabla)(b\partial_{a})^{2}\right]h^{(s)}(z;a)\qquad \nonumber\\
  &-&\frac{1}{L^{2}}\left[a^{2}(b\partial_{a})^{2}+(s-2)b^{2}
  +\frac{(ab)}{s}(s^{2}-6s+6)(b\partial_{a})\right]h^{(s)}(z;a)
\label{4.30}
\end{eqnarray}
Taking the trace we obtain the so-called gauge invariant Fronsdal operator
\begin{eqnarray}
  \frac{1}{2}\Box_{b}\Gamma^{(2,s)}(z;a,b)&=&\Box h^{(s)}(z;a)-(a\nabla)\left[(\nabla\partial_{a})
  -\frac{1}{2}(a\nabla)\Box_{a}\right]h^{(s)}(z;a)  \nonumber\\
  &-&\frac{1}{L^{2}}\left[a^{2}\Box_{a}+s^{2}+s(d-5)-2(d-2)\right]h^{(s)}(z;a),
\label{4.31}
\end{eqnarray}
which is the $AdS$ generalization of (\ref{1.3}).
This completes the discussion of the case $k=1$.

\subsection{The solutions for the cases $k>1$}
We return now to the general case of arbitrary $k$. After insertion of the solutions and semi-
solutions into (1.30),(3.23), we get for the gauge variation of the Christoffel symbol or the curvature
\begin{eqnarray}
&& L^{2k}\left\{\sum_{n=0}^{k} L^{-2n}\delta\Gamma_{n}^{(m,s)}\right\}
    = \frac{(-1)^{m}}{m!} L^{2k}(a\nabla)^{m+1} (b\partial_{a})^{m} \xi^{(s-1)}(z;a) \nonumber\\
    &&+ \sum_{r_1,r_2,r_3; r_3\not= k} \Delta_{r_1 r_2 r_3}\frac{(-1)^{l_{\max}}}{(l_{\max})!}
    (a^{2})^{r_1}(ab)^{r_2}(b^{2})^{r_3}(a\nabla)^{m-2k+1}(b\partial_{a})^{l_{\max}} \xi^{(s-1)}(z;a) ,\nonumber\\
    && (l_{\max} = m - r_2 -2r_3)
\label{4.32}
\end{eqnarray}
For the curvature the first term on the right and the first term in the sum for $(r_1r_2r_3) = (k,0,0)$ vanish since
for this one
\begin{equation}
l_{\max} = m = s
\label{4.33}
\end{equation}
The obstruction functions $\Delta_{r_1r_2r_3}$ can be expanded as a finite polynomial of $s-m$ with coefficients
no longer dependent on $s$
\begin{equation}
\Delta_{r_1r_2r_3} = \sum_{n\geq 0}(s-m)^{n}\Delta_{r_1r_2r_3}(n)
\label{4.34}
\end{equation}
(see (4.21),(4.22) for $k=1$ and Appendix B for $k=2$). In all the cases explicitly calculated we obtained
\begin{equation}
\Delta_{r_1r_2r_3}(0) = 0
\label{4.35}
\end{equation}
except for $r_1=k$ which for the gauge invariance requirement is irrelevant as we have just seen. We assume therefore that this is true for all k. This implies that the curvature
\begin{equation}
\Gamma^{(s)} = \sum_{k=0}^{[\frac{m}{2}]} \Gamma_{k}^{(s)}
\label{4.36}
\end{equation}
as expressed by (1.30),(3.23) after insertion of the $A$ is gauge invariant. This closes the curvature problem.

In the case of the Christoffel symbols $m<s$ we must apply a $b-$trace to (4.32). In the natural basis of functions this implies
\begin{eqnarray}
&&\frac{1}{2} \Box_{b} \frac{(-1)^{m-r_2-2r_3}}{(m-r_2-2r_3)!} (a^{2})^{r_1}(ab)^{r_2}(b^{2})^{r_3}(b\partial_{a})^{m-r_2-2r_3} \xi^{(s-1)}(z;a)=                    \nonumber\\
 &&\sum_{t_1t_2t_3} M_{t_1t_2t_3}^{r_1r_2r_3} \frac{(-1)^{m-t_2-2t_3}}{(m-t_2-2t_3-2)!}(a^{2})^{t_1}(ab)^{t_2}(b^{2})^{t_3}(b\partial_{a})^{m-t_2-2t_3 -2}\xi^{(s-1)}(z;a)
\label{4.37}
\end{eqnarray}
where the sum extends over
\begin{equation}
t_1 +  t_2 + t_3 = k-1
\label{4.38}
\end{equation}
Only in three cases the matrix $M$ has nonvanishing entries
\begin{eqnarray}
&(1)&\qquad t_1 =r_1 +1, t_2 = r_2 -2, t_3 =r_3: \nonumber\\
 && M_{r_1+1,r_2-2,r_3}^{r_1r_2r_3} = \frac{1}{2}r_{2}(r_2-1)\\
\label{4.39}
&(2)&\qquad t_1=r_1,t_2=r_2, t_3=r_3-1: \nonumber\\
 && M_{r_1,r_2,r_3-1}^{r_1r_2r_3} = r_3 [2(m-r_3-1) +d+1]\\
\label{4.40}
&(3)&\qquad t_1=r_1, t_2 =r_2-1, t_3 =r_3: \nonumber\\
 && M_{r_1,r_2-1,r_3}^{r_1r_2r_3} = r_2(m-s-r_2-2r_3)
\label{4.41}
\end{eqnarray}
This implies in particular that $M_{t_1t_2t_3}^{k00} = 0$.

Though an obstruction $\Delta_{00k}$ does not exist, we introduce now such function which is assumed to be expandable in powers of $s-m$ in the same way as the obstructions, for the only purpose to make the Christoffel symbols $b$-traceless, and insert it in (4.32). In this equation $r_3 = k$ is now allowed. Due to (4.37) $b$-tracelessness
of (4.32) amounts to
\begin{equation}
\sum_{r_{i}} M^{r_1r_2r_3}_{t_1t_2t_3}\Delta_{r_1r_2r_3} = 0
\label{4.42}
\end{equation}
For $k=1$ this implies
\begin{equation}
M^{010}_{000} \Delta_{010}+ M^{001}_{000}\Delta_{001} = 0
\label{4.43}
\end{equation}
and for $k=2$
\begin{equation}
M^{011}_{001} \Delta_{011} + M^{002}_{001}\Delta_{002} = 0
\label{4.44}
\end{equation}
\begin{equation}
M^{020}_{010} \Delta _{020} + M^{011}_{010}\Delta_{011} = 0
\label{4.45}
\end{equation}
\begin{equation}
M^{110}_{100} \Delta_{110} + M^{101}_{100}\Delta_{101} +M^{020}_{100}\Delta_{020} = 0
\label{4.46}
\end{equation}
Equations of the type (4.43), (4.44) are understood as definitions of the new functions $\Delta_{00k}$. For general
$k$ they are solved by
\begin{equation}
\Delta_{00k} = \frac{2k+s-m-1}{k[2(m-k)+d-1]}\Delta_{01,k-1}
\label{4.47}
\end{equation}
From this equation follows also that $\Delta_{00k}$ is of first order in $s-m$ since $\Delta_{01,k-1}$ is. For $k=1$ we used this equation already in (4.27) with $\Delta_{010}$ taken from (4.21). For $k=2$ we calculate $\Delta_{002}$ to all orders in $s-m$ in Appendix B.
Moreover there are for each $k$ a set of $\frac{1}{2}(k-1)(k+2)$ identities (for $k=2$ these are (4.45),(4.46)) which
must be fulfilled. From the results given in Appendix B it is easy to see that to the first order in $s-m$
these identities are satisfied. We assume therefore that all such identities are obeyed to all orders in $s-m$.

The term with factor $\Delta_{00k}$ which we want to add to (4.32) is a gauge variation (derivative) which we ought
to integrate to obtain a local linear functional acting on the gauge field $h^{(s)}$. An ansatz for this functional is
\begin{equation}
L^{-2k}\mathcal{L}_{k}[h^{(s)}] = \Delta_{00k} \left(\frac{b^{2}}{L^{2}}\right)^{k}\quad\sum_{l=0}^{m-2k}\frac{(-1)^{l}}{l!}(a\nabla)^{l}(b\nabla)^{m-2k-l}
(b\partial_{a})^{l} h^{(s)}(z;a)
\label{4.48}
\end{equation}
Then its gauge variation is analogous to the gauge variation of $\Gamma_{0}^{(m-2k,s)}$ (3.2)
\begin{eqnarray}
L^{-2k}\mathcal{L}_{k}[(a\nabla)\xi^{(s-1)}]&=& \Delta_{00k} \left(\frac{b^{2}}{L^{2}}\right)^{k} \langle \frac{(-1)^{m-2k}}{(m-2k)!}(a\nabla)^{m-2k+1}(b\partial_{a})^{m-2k} \xi^{(s-1)}(z;a) \nonumber\\
&+& \sum_{l=0}^{m-2k}\frac{(-1)^{l}}{l!}(a\nabla)^{l}[(b\nabla)^{m-2k-l}, (a\nabla)] (b\partial_{a})^{l}\xi^{(s-1)}(z;a)\rangle
\label{4.49}
\end{eqnarray}
The first sum on the right hand side is the object we want to integrate. The desired local linear functional of $h^{(s)}$ stands on the left hand side. But there are correction terms from the second sum of the right hand side which are higher order in $L^{-2}$ due to the commutator.  We assume that these correction term are included already in
the ansatz (1.30),(3.23) for the next level. Then we have to take account only of the modification
\begin{equation}
A_{002}^{(l)} \rightarrow A_{002}^{(l)}* = A_{002}^{(l)} +\Delta_{002}
\label{4.50}
\end{equation}
in equations such as (3.29).

The problem of the Christoffel symbols has therefore also a unique solution.

\section{The Riemann tensor}
\setcounter{equation}{0}
As explained in Section 1 we can derive the Riemann tensor (linearized in the HS field) by applying an antisymmetrizer to the de Wit-Freedman tensor. We introduce an antisymmetric tensor $A^{\mu\nu}$ from the tangential tensor space
(as a special case we could use $A^{\mu\nu} = a^{\mu}b^{\nu} -a^{\nu}b^{\mu}$) and contract it $s$ times with $R$.
Then we get for the flat case (up to a normalization)
\begin{equation}
R^{(s)}(z; A) = \prod_{i=1}^{(s)} (\partial_{\mu_{i}}A^{\mu_{i}\nu_{i}}) \quad h^{(s)}_{\nu_1,\nu_2...\nu_{s}}(z)
\label{5.1}
\end{equation}
We introduce the shorthand
\begin{equation}
\alpha^{\mu} = \partial_{\nu}A^{\nu\mu}
\label{5.2}
\end{equation}
so that (5.1) simplifies to
\begin{equation}
R^{(s)}(z; A) = h^{(s)}(z;\alpha)
\label{5.3}
\end{equation}
Here we always understand that vectorial differential operators contracted into $h^{(s)}$ act on the position $z$

The method of contracting $A^{\mu\nu}$ can be easily carried over to the AdS expressions (1.26), (1.30) with $m=s$.
We choose one $b^{\nu}$ and combine it with all possible $a^{\nu}$ and replace this pair by $A^{\mu\nu}$. This gives $s!$ terms for each term from $\Gamma_{k}^{(s)}$. We divide by $s!$ at the end. With the shorthands
\begin{eqnarray}
\alpha^{\mu} = \nabla_{\nu}A^{\nu\mu} \\
\label{5.4}
B^{\mu\nu} = A^{\mu}\quad_{\lambda} A^{\lambda\nu}
\label{5.5}
\end{eqnarray}
we obtain after some algebra
\begin{equation}
R^{(s)}(z; A) = \sum_{k=0}^{[s/2]}\frac{(-1)^{k}\Omega_{k}}{L^{2k}} h^{(s)}(z; \underbrace{\alpha,\alpha... \alpha}_{s-2k};
\underbrace{B, B... B}_{k})
\label{5.6}
\end{equation}
The function $h{(s)}$ depending on arguments $\alpha$ and $B$ must be read as follows
\begin{equation}
h^{(s)} (z; \underbrace{\alpha,\alpha...\alpha}_{s-2k};\underbrace{B,B ...B}_{k})
= \prod_{i=1}^{s-2k} \alpha^{\mu_{i}}\prod_{j=s-2k+1}^{s-k} B^{\mu_{j}\mu_{j+k}}\quad h^{(s)}_{\mu_1\mu_2...\mu_{s}}(z)
\label{5.7}
\end{equation}
We recognize that $R^{(s)}$ depends on a simple sequence of numerical constants $\Omega_{k}$ only. They can be expressed
in a surprisingly simple fashion by the coefficients $A_{r_1r_2r_3}^{(l)}$
\begin{equation}
\Omega_{k} = \sum_{r_1,r_2,r_3;r_1+r_2+r_3=k}\quad\sum_{l=2r_1+r_2}^{s-r_2-2r_3}A_{r_1r_2r_3}^{(l)}
\label{5.8}
\end{equation}
Inserting the coefficients $A_{r_1r_2r_3}^{(l)}$ at $s=m$ from (4.18)-(4.20) and Appendix B we get
\begin{eqnarray}
\Omega_{0}&=& 1 \\
\label{5.9}
\Omega_{1} &=& 4{s+1 \choose 4} +3{s+1 \choose 3} \\
\label{5.10}
\Omega_{2} &=& 70 {s+1 \choose 7} + 114 {s+1\choose 6} + 45 {s+1\choose 5}
\label{5.11}
\end{eqnarray}
It is remarkable that these coefficients $\Omega_{k}$ vanish when $2k > s$ as they should. Some details of the
arguments leading to (5.8) - (5.11) are presented in Appendix C.

\section{Conclusions}
\setcounter{equation}{0}
We derived de Wit-Freedman formulas for the curvature and the generalized Christoffel symbols both linearized in the higher spin field $h^{(s)}(z;a)$ on $AdS_{d+1}$ space to the second order ("level")in the expansion in powers of the $AdS$ inverse radius squared. These expressions are complete for spin smaller or equal five. The Riemann curvature
is then derived in the same order, it permits to obtain Ricci and Weyl tensors by elementary manipulations.
The concepts and arguments given here allow to formulate a computer algorithm by which higher orders necessary for spin greater or equal six can be obtained. New obstacles are not expected.

\subsection*{Acknowledgements}
\quad This work is supported by the German
Volkswagenstiftung. The work of R.~M. was
supported in part by the INTAS grant \#03-51-6346.

\section*{Appendix A: Derivation of $c_{s}$}
\setcounter{equation}{0}
\renewcommand{\theequation}{A.\arabic{equation}}
For simplicity we consider the flat space with dimension not smaller than two.
Set $\mu_{i} =1, \nu_{i} =2$ for all $i$ in (1.6). Then instead of (1.18)
with still undetermined $c_{s}$ we have by expansion
\begin{equation}
R_{12,12,12,...12}^{(s)} = c_{s}^{-1}\sum_{l=0}^{s}(-1)^{s-l}{s \choose l}
\Gamma_{1,1...1_{l},2,2,...2_{s};1,1...1_{s-l},2,2...2_{s}}^{(s)}
\label{6.1}
\end{equation}
In turn from (1.10) we have
\begin{equation}
\Gamma_{1,1...1_{l},2,2...2_{s};1,1...1_{s-l},2,2...2_{s}}^{(s)} =
(s!)^{-1}\sum_{P(\nu)} R_{1P(\nu_1),1P(\nu_2)...1P(\nu_{l});2P(\nu_{l+1}),
2P(\nu_{l+2})...2P(\nu_{s})}^{(s)}
\label{6.2}
\end{equation}
where $P(\nu_{i})$ denote the permutations of the $\nu_{i}$ and over all these we sum in (\ref{6.2}). But only those permutations count by symmetry for which
\begin{eqnarray}
P(\nu_{i}) &=& 2 \qquad \textnormal{whenever} \qquad 1\leq i \leq l \\
P(\nu_{i}) &=& 1 \qquad \textnormal{whenever} \qquad l+1 \leq 1 \leq s
\label{6.3}
\end{eqnarray}
Counting the number of permutations that are relevant we obtain $l!(s-l)!$ so that
\begin{equation}
\Gamma_{11,1...1_{l},2,2...2_{s};1,1...1_{s-l},2,2..._{s}}^{(s)} =
(-1)^{s-l}{s \choose l}^{-1} R_{12,12...12}^{(s)}
\label{6.4}
\end{equation}
It follows that $c_{s} = 1$ as claimed in (\ref{1.16}).

\section{Appendix B: Results for $k=2$}
\setcounter{equation}{0}
\renewcommand{\theequation}{B.\arabic{equation}}
For the curvature $m=s$ the term $\{r_1r_2r_3\} = \{k00\}$ and the first term of (4.32) drop out since the $s$-th derivative with respect to $a$ of $\xi^{(s-1)}$ vanishes. The remaining obstructions reduce to $\Delta_{r_1r_2r_3}(0)$ and are calculated for $k=2$ and shown to be zero (see below). Assuming that this holds also for all levels $k$, the derivation of the de Wit-Freedman curvature is completed.
We perform polynomial expansions of $A$ and the integration constants $c$ in powers of $s-m$
\begin{equation}
A^{(l)}_{r_1r_2r_3} = \sum_{n\geq 0} (s-m)^{n} A^{(l)}_{r_1r_2r_3}(n)
\label{7.1}
\end{equation}
\begin{equation}
c_{r_1r_2r_3} = \sum_{n\geq 0} (s-m)^{n} c_{r_1r_2r_3}(n)
\label{7.2}
\end{equation}
We use the notation of the "inverse Pochhammer"
\begin{equation}
[x]_{n} = {x \choose n} n!
\label{7.3}
\end{equation}
Then we obtain for the all (four) cases $\{r_1r_2r_3\}$ with exception of $r_1 =2$ and $r_2=2$:
\begin{equation}
c_{r_1r_2r_3}(0) = \Delta_{r_1r_2r_3} (0) = 0
\label{7.4}
\end{equation}
We give the coefficients $A_{011}^{(l)}$ to all orders in $s-m$ but the other ones only to the first order,
remember $\gamma_{m} =(d+2m-3)^{-1}$
\begin{equation}
A_{011}^{(l)}(0) = -\frac{1}{12}[m-l+1]_{6} + \frac{1}{24}(2m-11)[m-l+1]_{5} +\frac{1}{8}(m-3)[m-l+1]_{4}
\label{7.5}
\end{equation}
\begin{eqnarray}
A_{011}^{(l)}(1) &=&  -\frac{1}{3}[m-l+1]_{5} +\frac{1}{4}(m-4)[m-l+1]_{4} +\frac{1}{12}[m+1]_{2}[m-l]_{3}
   \nonumber\\               && +\frac{1}{4}\gamma_{m}[m+1]_{2}\{-[m-l]_{3}+(m-2)[m-l-1]_{2}\}+ c_{011}(1)
\label{7.6}
\end{eqnarray}
\begin{equation}
c_{011}(1) = \Delta_{011}(1) = \frac{1}{2}\gamma_{m} [m+1]_{4}
\label{7.7}
\end{equation}
\begin{equation}
A_{011}^{(l)}(2) =  -\frac{1}{4}[m-l+1]_{4} +\frac{1}{4}[m+1]_{2}[m-l-1]_{2}
\label{7.8}
\end{equation}
\begin{equation}
c_{011}(2) = \Delta_{011}(2) = 0
\label{7.9}
\end{equation}
\begin{equation}
A_{101}^{(l)}(0) = \frac{1}{18} [m-l+2]_{6} -\frac{1}{12}(m-2)[m-l+2]_{5} +\frac{1}{36}[m+2]_{3}[m-l]_{3}
\label{7.10}
\end{equation}
\begin{eqnarray}
A_{101}^{(l)}(1) &=& \frac{7}{45}[m-l+2]_{5} -\frac{1}{12}(3m-4)[m-l+2]_{4} +\frac{1}{12}[m+2]_{3}[m-l]_{2}
\nonumber\\ &&+\frac{1}{12}\gamma_{m}[m+1]_{2}\{2[m-l]_{3} -3(m-2)[m-l]_{2}\} + c_{101}(1)
\label{7.11}
\end{eqnarray}
\begin{equation}
c_{101}(1) = \Delta_{101}(1) = \frac{1}{12}\gamma_{m} m[m+1]_{4}
\label{7.12}
\end{equation}
\begin{equation}
A_{020}^{(l)}(0) = \frac{1}{8} [m-l+2]_{6} -\frac{1}{4}(m-3)[m-l+2]_{5} +\frac{1}{8}(m-2)(m-3)[m-l+2]_{4}
\label{7.13}
\end{equation}
\begin{eqnarray}
A_{020}^{(l)}(1) &=& \frac{1}{4}[m-l+2]_{5} -\frac{1}{8}(2m-5)[m-l+2]_{4} -\frac{1}{4}[m+1]_{2}\{[m-l]_{3}
\nonumber\\ && -(m-3)[m-l]_{2}\} + c_{020}(1)
\label{7.14}
\end{eqnarray}
\begin{equation}
c_{020}(1) = \Delta_{020}(1) = \frac{1}{8} [m+1]_{4}
\label{7.15}
\end{equation}
\begin{eqnarray}
A_{110}^{(l)}(0) &=& -\frac{1}{6}[m-l+3]_{6} +\frac{1}{12}(5m-8)[m-l+3]_{5} -\frac{1}{12}(m-1)(3m-4)[m-l+3]_{4}
\nonumber\\ &&-\frac{1}{12}[m+2]_{3}\{[m-l+1]_{3} -(m-1)[m-l+1]_{2}\}
\label{7.16}
\end{eqnarray}
\begin{eqnarray}
A_{110}^{(l)}(1) &=& -\frac{1}{6}[m-l+3]_{5} +\frac{1}{24}(6m-5)[m-l+3]_{4} +\frac{1}{6}[m+1]_{2}[m-l+1]_{3}
\nonumber\\ &&-\frac{1}{12}(4m-1)[m+1]_{2}[m-l+1]_{2} + c_{110}(1)
\label{7.17}
\end{eqnarray}
\begin{equation}
c_{110}(1) = \Delta_{110}(1) = \frac{1}{24}(2m+3)[m+1]_{4}
\label{7.18}
\end{equation}
Whereas by definition
\begin{eqnarray}
A_{r_1r_2r_3}^{(l_{max}+1)}(n) = A_{r_1r_2r_3}^{(m+1-r_2-2r_3)}(n) = \Delta_{r_1r_2r_3}(n) \\
\label{7.19}
A_{r_1r_2r_3}^{(l_{min}-1)}(n) = A_{r_1r_2r_3}^{(2r_1+r_2-1)}(n) = 0
\label{7.20}
\end{eqnarray}
and the latter equation determines $c_{r_1r_2r_3}(n)$, the equality of $\Delta_{r_1r_2r_3}$ and $c_{r_1r_2r_3}$ in all cases listed here is a nontrivial result.

Now we study the exceptional cases $\{200\}$ and $\{002\}$. There is no dependence on $s$ in the first case.
\begin{eqnarray}
A_{200}^{(l)} &=& \frac{1}{18} [l]_{6} -\frac{1}{30}(5m-6)[l]_{5} +\frac{1}{8} m(m-2)[l]_{4}\\
\label{7.21}
\Delta_{200} &=& \frac{1}{360}(5m+12)[m+2]_{5}
\label{7.22}
\end{eqnarray}
The second case gives to all orders in $s-m$
\begin{equation}
A_{002}^{(l)}(0) = \frac{1}{360}\{5[m-l+1]_{6} -3[m-l+1]_{5} + 15[m-l]_{4}\}
\label{7.23}
\end{equation}
\begin{equation}
A_{002}^{(l)}(1) = \frac{1}{12}\left[[m-l+1]_{5} -2[m-l]_{4} + \gamma_{m}[m+1]_{2}\{[m-l-1]_{3} +3[m-l-2]_{2}\}\right]
\label{7.24}
\end{equation}
\begin{equation}
A_{002}^{(l)}(2) = \frac{1}{8} [m-l]_{4} +\frac{1}{12}\gamma_{m}[m+1]_{2}\{[m-l-1]_{3} +6 [m-l-2]_{2}\}
\label{7.25}
\end{equation}
\begin{equation}
A_{002}^{(l)}(3) = \frac{1}{4} \gamma_{m}[m+1]_{2}[m-l-2]_{2}
\label{7.26}
\end{equation}
There is no proper obstruction in this case. But the trace compensator $\Delta_{002}$ is from (4.47) and (B.7),(B.9)
\begin{equation}
\Delta_{002} = \frac{1}{4}(s-m)(s-m+3)\gamma_{m}\gamma_{m-1}[m+1]_{4}
\label{7.27}
\end{equation}

\section{Appendix C: Derivation of the Riemann tensor}
\setcounter{equation}{0}
\renewcommand{\theequation}{C.\arabic{equation}}
By the total symmetry of the tensor field $h^{(s)}$
\begin{equation}
(b\partial_{a})^{l}h^{(s)}(z; a) = [s]_{l}h^{(s)}(z; b,a)
\label{C.1}
\end{equation}
where
\begin{equation}
[s]_{l} = l!{s \choose l}
\label{C.2}
\end{equation}
it is impossible to contract a pair $a, b$ in its arguments to the antisymmetric tensor $A$. On the other hand
the number of $a$ (resp. $b$) in the arguments of $h^{(s)}$ is the same as the number of $b$ (resp. $a$) in the prefactors
\begin{equation}
(a^{2})^{r_1}(ab)^{r_2}(b^{2})^{r_3}(a\nabla)^{l-2r_1-r_2}(b\nabla)^{s-l-r_2-2r_3}
\label{C.3}
\end{equation}
Therefore it is also impossible to contract a pair $a, b$ from inside the prefactors (C.3) to an
$A$.

Contracting an $a$ from $(a\nabla)^{l-2r_1-r_2}$ with a $b$ from $h^{(s)}$ yields
\begin{equation}
(a^{\mu}\nabla_{\mu})b^{\lambda} \rightarrow -\nabla_{\mu}A^{\mu\lambda} = - \alpha^{\lambda}
\label{C.4}
\end{equation}
and contracting a $b$ from $(b\nabla)^{s-l-r_2-2r_3}$ with an $a$ from $h^{(s)}$ we get
\begin{equation}
(b^{\mu}\nabla_{\mu})a^{\lambda} \rightarrow +\nabla_{\mu}A^{\mu\lambda} = +\alpha^{\lambda}
\label{C.5}
\end{equation}
The combinatorical factor and sign is
\begin{equation}
(-1)^{l} \frac{l!(s-l)!}{(2r_1+r_2)!(2r_3+r_2)!}
\label{C.6}
\end{equation}
After this contraction there are $2r_1 +r_2$ $b$s and $r_2+2r_3$ $a$s left as arguments in $h^{(s)}$.

Next we make the contraction with $(ab)^{r_2}$. We take one pair $g_{\mu\nu}a^{\mu}b^{\nu}$ and contract
it with $b^{\lambda}a^{\kappa}$ from the arguments of $h^{(s)}$. This yields
\begin{equation}
g_{\mu\nu}(-A^{\mu\lambda})(+A^{\nu\kappa}) = B^{\lambda\kappa}
\label{C.7}
\end{equation}
There is no sign factor and the combinatorical factor is
\begin{equation}
\frac{(2r_1+r_2)!(2r_3+r_2)!}{(2r_1)!(2r_3)!}
\label{C.8}
\end{equation}
After these contractions the number of $a$s left in $h^{(s)}$ is $2r_3$ and the number of $b$s is $2r_1$.

Now we contract with $a$s from $(a^{2})^{r_1}$. We get
\begin{equation}
g_{\mu\nu}a^{\mu}a^{\nu}b^{\lambda} b^{\kappa} = -B^{\lambda\kappa}
\label{C.9}
\end{equation}
The sign and combinatorical factor is
\begin{equation}
(-1)^{r_1}(2r_1)!
\label{C.10}
\end{equation}
In the same fashion we obtain for the contraction of $(b^{2})^{r_3}$ the sign and combinatorical factor
\begin{equation}
(-1)^{r_3}(2r_3)!
\label{C.11}
\end{equation}
Thus we finished the contractions.

We collect all sign and combinatorical factors and obtain
\begin{equation}
(-1)^{l}{s\choose l}(-1)^{l-2r_1-r_2}l!(s-l)!(-1)^{r_1+r_3} = s!(-1)^{r_1+r_2+r_3} = s! (-1)^{k}
\label{C.12}
\end{equation}
In this way the proof of the results of Section 5 is almost completed.

The technical problem to perform the summation (5.8) is solved by using \cite{GR} and manipulating the summation label
in such a fashion that the first term in each sum is one as it is in the quoted formula.

\end{document}